\documentclass[english,aps, twocolumn]{revtex4}
\usepackage[T1]{fontenc}
\usepackage[latin9]{inputenc}
\usepackage{bm}
\usepackage{amssymb}
\usepackage{esint}

\usepackage{babel}

\begin{document}

\title{Exciton Dynamics in Photosynthetic Complexes: Excitation by Coherent
and Incoherent Light }

\author{Tom\'{a}\v{s} Man\v{c}al$^{1}$ and Leonas Valkunas$^{2,3}$}

\affiliation{$^{1}$Charles University in Prague, Faculty of Mathematics and Physics,
Ke Karlovu 5, CZ-121 16 Prague 2, Czech Republic}

\affiliation{$^{2}$Institute of Physics, Savanoriu Avenue 231, 02300 Vilnius,
Lithuania}

\affiliation{$^{3}$Department of Theoretical Physics, Faculty of Physics of Vilnius
University, Sauletekio Avenue 9, build. 3, 10222 Vilnius Lithuania}
\begin{abstract}
In this paper we consider dynamics of a molecular system subjected
to external pumping by a light source. Within a completely quantum
mechanical treatment, we derive a general formula, which enables to
asses effects of different light properties on the photo-induced dynamics
of a molecular system. We show that once the properties of light are
known in terms of certain two-point correlation function, the only
information needed to reconstruct the system dynamics is the reduced
evolution superoperator. The later quantity is in principle accessible
through ultrafast non-linear spectroscopy. Considering a direct excitation
of a small molecular antenna by incoherent light we find that excitation
of coherences is possible due to overlap of homogeneous line shapes
associated with different excitonic states. In Markov and secular
approximations, the amount of coherence is significant only under
fast relaxation, and both the populations and coherences between exciton
states become static at long time. We also study the case when the
excitation of a photosynthetic complex is mediated by a mesoscopic
system. We find that such case can be treated by the same formalism
with a special correlation function characterizing ultrafast fluctuations
of the mesoscopic system. We discuss bacterial chlorosom as an example
of such a mesoscopic mediator and propose that the properties of energy
transferring chromophore-protein complexes might be specially tuned
for the fluctuation properties of their associated antennae.
\end{abstract}
\maketitle

\section{Introduction }

In recent years, primary processes in photosynthesis have received
a renewed interest from a broader physical community thanks to experimental
observation of coherent energy transfer in some photosynthetic systems.
The ground breaking coherent two-dimensional electronic spectroscopy
(2D-ES) experiment of Engel \emph{et al.} \cite{Engel2007a} has led
to new appreciations of the role that may be played by coherent dynamics
in excitation energy transfer (EET), and of the quantum mechanical
nature of photosynthetic systems in general \cite{Scholes2010a}.
Special theoretical effort has been made to understand the role of
noise \cite{Darius2004,Plenio2008,Castro2008,Rebentrost2009,Mohseni2009}
in the dynamics of excitation energy transfer, and the role of coherence
\cite{Ishizaki1,Ishizaki2,Ishizaki-PNAS,Cheng-Review,Olsina2010a}
in excitonicaly coupled systems. On the experimental front, the method
of coherent 2D-ES \cite{Jonas2003a,Cho2008a} has established itself
as a tool opening new window into the details of energy transfer dynamics
in photosynthetic \cite{Brixner2005a,Cho2005a,Calhoun2009,Cohen2009,Ginsberg2009,Sanda2009},
and other molecular systems \cite{Milota2009,Darius2004,Collini2009}.
Coherent effects have been now reported in different molecular systems,
often biologically relevant \cite{Lee2007a,Collini2009} - a generality
that asks for a search of the possible evolutionary advantage underlying
their abundance in photosynthetic pigment-protein complexes. 

The principle pigment molecules responsible for the primary processes
of photosynthesis are chlorophylls (Chls) and bacteriochlorophylls
(BChls) \cite{Blankenship,Chls}. They are involved in accumulation
of light energy via the excitation energy transfer to specific pigment-protein
complexes - reaction centers. Spectral variability of photosynthetic
light-harvesting pigment-protein complexes arises either from excitonic
interactions between pigment molecules or from their interactions
with protein surrounding. Both these interactions are the main factors
determining the excitation dynamics in light-harvesting \cite{AVG}.
Excitonic aggregates are subject to interaction with two types of
environments, and they provide means of transferring energy from one
environment to another. First of these environments, the radiation,
is under natural conditions at much higher temperature than the second
environment, the protein scaffold and indeed the photosynthetic chemical
machinery as a whole. The excess of photons on suitable wavelength
in the radiational environment is used to excite spatially extended
antenna systems that concentrate excitation energy to the reaction
center, which in turn drives charge transfer processes across cellular
membranes to create the transmembrane potential and the pH gradient
\cite{Blankenship}. 

Non-equilibrium processes occurring in photosynthetic systems during
light harvesting are conveniently described by reduced density matrix
(RDM) theory \cite{AVG,MayKuehnBook,Bruggemann07} which has an advantage
of being applicable to disordered statistical ensembles that the experiments
often deal with. However, with recent 2D experiments that enable us
to distinguish the homogeneous and inhomogeneous spectral broadening,
and with the progress in single molecular spectroscopy \cite{SingleBook}
we can gain insight into the time evolution characteristic to single
molecules interacting with their environment \cite{Valkunas2007,Janusonis2008}.
This fact enables us to return to the wavefunction formalism and to
look at light harvesting from the point of view which takes the superposition
principle of quantum mechanics seriously. It has been show that such
an approach yields many interesting insights into the emergence of
the classical properties of molecular system from their underlying
quantum mechanical nature \cite{SchlosshauerBook,DecoherenceBook}.
As the light-harvesting processes seem to operate on the interface
between classical and quantum worlds it seems appropriate to look
at them from the point of view of the decoherence program of Zeh,
Zurek and others \cite{Zeh70,Zurek82}.

The process of light harvesting could then be describes as follows.
First, the system is in an {}``equilibrium'' initial state $|\Psi_{0}\rangle$
characterized by the excitonic ground state $|g\rangle$, the state
of protein (phonon) environment $|\Phi_{P}\rangle$ corresponding
to this electronic ground state and some state of light $|\Xi_{0}\rangle$,
i.e. \begin{equation}
|\Psi_{0}\rangle=|g\rangle|\Phi_{B}\rangle|\Xi_{0}\rangle.\label{eq:state_ini}\end{equation}
The light-harvesting occurs when the state of light is such that the
time evolution of the system leads to population of higher excited
states $|e_{n}\rangle$ of photosynthetic antenna. These states are
formed from excited states of Chls and other chromophores, such as
carotenoids \cite{AVG}. We denote these combined excited states as
excitons. In the first approximation, photosynthetic antenna remains
in the excited state until the excitation energy is transferred to
the reaction center. This happens much faster than competing process
of spontaneous emission which can therefore be neglected in our discussion.
When the interaction of the antenna with light is switched on, the
change occurring in the ground state portion of the total state vector
after the passage of time $\Delta t$ is\begin{equation}
|\Psi_{0}\rangle\rightarrow\alpha_{\Delta t}|g\rangle|\Phi_{B}\rangle|\Xi_{0}\rangle+\sum_{n}\beta_{\Delta t}^{(n)}|e_{n}\rangle|\Phi_{B}\rangle|\Xi^{\prime}\rangle\label{eq:state_exc}\end{equation}
The subsequent time evolution of the excited state portion of the
state vector is independent of the ground state part, and we can thus
look at it separately. Because we neglect spontaneous emission, any
excitation to higher excited state, as well as transitions between
exciton states due to the light, the state vector $|\Xi^{\prime}\rangle$
remains approximately unentangled with excitons and the protein bath
for the rest of the energy transfer process. It can therefore be omitted.
The initial state for the energy transfer process thus reads\begin{equation}
|\Psi_{e}(t_{0})\rangle=\sum_{n}\beta^{(n)}(t_{0})|e_{n}\rangle|\Phi_{B}\rangle,\label{eq:ini_in_ex}\end{equation}
where we omitted the lower index $\Delta t$. If the basis of the
states $|e_{n}\rangle$ is chosen so that the molecular Hamiltonian
is diagonal, the energy transfer occurs only due to interaction of
excitons with their surrounding environment. This interaction leads
to an entanglement of excitons and the environment \begin{equation}
|\Psi_{e}(t)\rangle=\sum_{n}\beta^{(n)}(t)|e_{n}\rangle|\Phi_{B}^{(n)}(t)\rangle.\label{eq:entanglement}\end{equation}
After a sufficiently long time the environment state vectors corresponding
to different electronic state diverge maximally and the reduced density
matrix becomes diagonal in some basis, i.e. \[
\rho(t)=tr_{B}\{|\Psi_{e}(t)\rangle\langle\Psi_{e}(t)|\}\]
\begin{equation}
=\sum_{mn}\beta^{(n)}(t)(\beta^{(m)}(t))^{*}\langle\Phi_{B}^{(m)}(t)|\Phi_{B}^{(n)}(t)\rangle.\label{eq:dens_mat_from_ex}\end{equation}
Often, to a good approximation, such preferred basis is the one in
which the electronic Hamiltonian is diagonal, the so-called excitonic
basis. However, notable corrections to this rule are predicted even
for weak system-bath coupling \cite{Tannor,Olsina2010a}.

The final state of the energy transfer is the one in which just reaction
centers are populated\begin{equation}
|\Psi_{e}(t)\rangle=\sum_{k}\beta^{(RC_{k})}(t)|e_{RC_{k}}\rangle|\Phi_{B}^{(RC_{k})}(t)\rangle.\label{eq:in_RC}\end{equation}
The last step of the energy transfer, from the antenna to the reaction
center is often slower than typical transfer times between antenna
complexes, and so the final state is well localized on the reaction
center, and coherences between individual reaction centers are unlikely
to survive. 

It is clear from the above discussion, that decoherence during the
energy transfer in the antenna is determined by the evolution of the
environmental degrees of freedom (DOF). The decoherence from the rest
of the system might be required for the localization of the energy
in the reaction center, but there is no obvious reason for fast decoherence
during the initial steps of energy transfer in the antenna, apart
from the fact that a bath formed by a completely random disordered
environment would lead to just such fast decoherence. It has been
suggested before that the protein environment might play a more active
role in steering and protecting electronic excitation \cite{Engel2007a,Lee2007a}
and controlling the decoherence might be one of the possible pathways
to more robust EET. 

There is however one important caveat in the above scheme. The initial
condition, Eq. (\ref{eq:ini_in_ex}), has been introduced artificially
into Eq. (\ref{eq:state_exc}) as a result of an interaction occurring
during some short time interval $\Delta t$. If the system is continuously
pumped, individual contributions similar to Eq. (\ref{eq:ini_in_ex})
will interfere, possibly disabling any effect of cooperative involvement
of the bath. It is even more important to consider the question what
are the effects of natural sun light \cite{Brumer}, i.e. whether
the coherent scenario outlined above is plausible for the photosynthetic
system \emph{in vivo,} or not. This depends strongly on the nature
of the excitation process, whether it occurs in discrete independent
jumps of the kind described by Eq. (\ref{eq:state_exc}), or continuously
over a long period of uncertainty interval of the photon arrival.
The former view is usually held in support of the relevance of ultrafast
spectroscopic finding for \emph{in vivo} function of the photosynthetic
systems \cite{Cheng-Review}. Below we derive a general formula which
enables us to describe all these regimes by a unified formalism, and
also enables us to place the observables of ultrafast coherent spectroscopy
in perspective with the dynamics under natural conditions. In a somewhat
extended form our result is also applied to another case cited in
support of the utility of coherent dynamics in photosynthetic systems,
a case where a small photosynthetic complex is excited through another,
possibly mesoscopic, antenna \cite{Cheng-Review}. 

The paper is organized as follows. Next section introduces a rather
general model of photosynthetic aggregate. In Section \ref{sec:Excitation-by-Coherent}
we discuss the dynamics of a system excited by coherent pulsed light
and the observables of the ultra-fast non-linear spectroscopy. Section
\ref{sec:Excitation-by-Thermal} is concerned with the excitation
of a photosynthetic system by the light from a general source. Implications
of the theory for excitation by thermal and coherent light, as well
as excitation mediated by mesoscopic system are discussed in Section
\ref{sec:Discussion}.

\section{Hamiltonian of a Model Photosynthetic System\label{sec:Hamiltonian-of}}

In this section we briefly review the excitonic model that was very
successfully applied to model the spectroscopic properties of Chl-
and Bchl- based light harvesting chromophore--protein complexes (see
e.g. \cite{Cho2005a}). We assume $N$ monomers with ground states
$|g_{n}\rangle$, excited states $|\tilde{e}_{n}\rangle$, $n=1,\dots,N$,
and with electronic transition energies $\tilde{\varepsilon}_{n}$.
These monomers are interacting with the phonon bath of protein DOF
so that the Hamiltonian of the monomer reads\begin{equation}
H_{n}=(T+V_{g})|g_{n}\rangle\langle g_{n}|+(\tilde{\varepsilon}_{n}+T+V_{e})|\tilde{e}_{n}\rangle\langle\tilde{e}_{n}|.\label{eq:mono-ham}\end{equation}
Here, $T$ is the kinetic energy operator of the bath, and $V_{g}$
and $V_{e}$ are the potential energy operators of the bath when the
system is in the electronic ground- and excited states, respectively.
We set the ground state electronic energy to zero for conveniency.
The Hamiltonian, Eq. (\ref{eq:mono-ham}) can be split into the pure
bath, pure electronic and the interaction terms so that\[
H_{n}=\underbrace{(T+V_{g})\otimes\mathbb{I}_{M}^{n}}_{H_{B}^{n}}+\underbrace{\mathbb{I}_{B}^{n}\otimes(\tilde{\varepsilon}_{n}+\langle V_{e}-V_{g}\rangle_{eq})|\tilde{e}_{n}\rangle\langle\tilde{e}_{n}|}_{H_{M}^{n}}\]
\begin{equation}
+\underbrace{(V_{e}-V_{g}-\langle V_{e}-V_{g}\rangle_{eq})\otimes|\tilde{e}_{n}\rangle\langle\tilde{e}_{n}|}_{H_{M-B}^{n}}.\label{eq:Ham_n_splitted}\end{equation}
Here, $\mathbb{I}_{B}$ is the unity operator on the bath Hilbert
space and $\mathbb{I}_{M}$ is the unity operator on the Hilbert space
of the electronic states. The equlibrium average $\langle V_{e}-V_{g}\rangle_{eq}$
of the potential energy operators was added to the electronic energy
so that the interaction term is zero for the system in equilibrium.

In chromophore--protein complexes many such monomers are coupled by
resonance coupling. The whole complex can be described by means of
collective states including the ground state\begin{equation}
|g\rangle=\prod_{n=1}^{N}\otimes|g_{n}\rangle,\label{eq:ground}\end{equation}
one excitation states\begin{equation}
|\bar{e}_{a}\rangle=\prod_{n=1}^{a-1}\otimes|g_{n}\rangle\otimes|\tilde{e}_{a}\rangle\otimes\prod_{m=a+1}^{N}\otimes|g_{m}\rangle,\label{eq:oneexc}\end{equation}
and states containing higher number of excitations. For the sake of
brevity we now stop writing the symbol of the direct product $\otimes$
and the unity operators $\mathbb{I}_{B}^{n}$ etc. explicitely. The
total Hamiltonian of the complex including resonance interaction is
then defined as\[
H_{B}+H_{M}+H_{M-B}=\sum_{n=1}^{N}H_{B}^{n}+\sum_{n=1}^{n}H_{M}^{n}\]
\begin{equation}
+\sum_{n\neq m}^{N}J_{nm}|\bar{e}_{n}\rangle\langle\bar{e}_{m}|+\sum_{n=1}^{n}H_{M-B}^{n}.\label{eq:Htot}\end{equation}
If the system-bath interaction with bath is weak, the referred basis
into which the electronic system relaxes due to interaction with the
bath is, to a good approximation, the one in which the electronic
part of the Hamiltonian is diagonal. Let us denote these states as
$|e_{n}\rangle$. They are usually termed excitons and they represent
certain linear combination of the collective states $|\bar{e}_{n}\rangle$
where excitations are localized on individual chromophore molecules.
One of the most important characteristics of this model is that it
does not include direct relaxation of the electronic excited states
to the ground states due to electron-phonon coupling. This is well
satisfied by Chls and BChls on the ultrafast time scale of which light
harvesting processes occur.

\section{Excitation by Coherent Pulsed Light and Non-Linear Spectroscopy\label{sec:Excitation-by-Coherent}}

Let us now consider experimental methods which provide information
about time evolution of excited states of photosynthetic systems.
Because of the timescale of EET processes, spectroscopy with ultrashort
time resolution is a necessary tool. The interaction of the pulsed
coherent light with the photosynthetic system is well described in
semi-classical approximation \cite{MukamelBook}. Electric field of
the light is then considered as an external parameter of the system
Hamiltonian. Electronic DOF can be prepared very fast in an excited
state, not affecting, to a good approximation, the bath DOF. Thus,
in an experiment with an ideal time resolution, we would have the
system prepared in the excited state, Eq. (\ref{eq:ini_in_ex}). The
time evolution of the system is governed by the Schr\"{o}dinger equation\[
\frac{\partial}{\partial t}|\Psi_{e}(t)\rangle=-\frac{i}{\hbar}(H_{B}+H_{M}+H_{M-B})|\Psi_{e}(t)\rangle\]
\begin{equation}
+\delta(t)|\Psi_{e}(t_{0})\rangle,\label{eq:with_delta}\end{equation}
with initial condition $|\Psi_{e}(t)\rangle=0$ for $t<t_{0}$. The
last term in Eq. (\ref{eq:with_delta}) describes the ultrafast event
of the molecule--radiation interaction. Formal solution of this equation
reads $|\Psi_{e}(t)\rangle=U_{B}(t)U_{M}(t)U_{M-B}(t)|\psi_{e}(t_{0})\rangle$,
where we defined evolution operators $U_{B}(t)$, $U_{M}(t)$ of the
bath and the molecule, respectively, as \begin{equation}
U_{B}(t)=\Theta(t-t_{0})\exp\{-\frac{i}{\hbar}H_{B}(t-t_{0})\},\label{eq:evol_op0}\end{equation}
\begin{equation}
U_{M}(t)=\Theta(t-t_{0})\exp\{-\frac{i}{\hbar}H_{M}(t-t_{0})\},\label{eq:evol_op}\end{equation}
and the remaining interaction evolution operator as\[
U_{M-B}(t)=\Theta(t-t_{0})\exp\{-\frac{i}{\hbar}\int\limits _{t_{0}}^{t}d\tau U_{B}^{\dagger}(\tau)U_{M}^{\dagger}(\tau)\]
\begin{equation}
\times H_{M-B}U_{M}(\tau)U_{B}(\tau)\},\label{eq:sol_with_delta}\end{equation}
After excitation, the process of energy transfer proceeds according
to the description presented in Introduction and can be experimentally
monitored.

\subsection{Evolution superoperator}

Matrix elements of the RDM of the molecule, which holds the information
about the population probabilities and the amount of coherence between
electronic states are given by expectation value of projectors $|e_{n}\rangle\langle e_{m}|$,
\[
\rho_{nm}(t)=\langle\psi_{e}(t)|e_{m}\rangle\langle e_{n}|\psi_{e}(t)\rangle\]
\[
=tr_{B}\{\langle e_{n}|\psi_{e}(t)\rangle\langle\psi_{e}(t)|e_{m}\rangle\}\]
\[
=\langle e_{n}|tr_{B}\{U_{M}(t)U_{M-B}(t)\underbrace{\sum_{ab}\beta^{(a)}(\beta^{(b)})^{*}|a\rangle\langle b|}_{\rho_{0}}\]
\begin{equation}
\times\underbrace{|\Phi_{B}\rangle\langle\Phi_{B}|}_{W_{eq}}U_{M-B}^{\dagger}(t)U_{M}^{\dagger}(t)\}|e_{m}\rangle\label{eq:P_nmt}\end{equation}
This can be rewritten by defining an evolution superoperator ${\cal U}(t)$
which acts on initial density matrix $\rho_{0}W_{eq}$, i.e.\[
\rho_{nm}(t)=tr_{B}\{\langle e_{n}|W(t)|e_{m}\rangle\}\]

\begin{equation}
=\langle e_{n}|{\cal U}^{(e)}(t)\rho_{0}W_{eq}|e_{m}\rangle,\label{eq:P_nm_sup}\end{equation}
The matrix elements of the superoperator read \[
{\cal U}_{abcd}^{(e)}(t)=\langle a|U_{M}(t)U_{M-B}(t)|c\rangle\dots\]
\begin{equation}
\times\langle d|U_{M-B}^{\dagger}(t)U_{M}^{\dagger}(t)|b\rangle,\label{eq:sup_elements}\end{equation}
where the dots $\dots$ denote where an operator on which ${\cal U}^{(e)}(t)$
acts has to be inserted. The reduced evolution operator $\bar{{\cal U}}^{(e)}(t)$
defined as\begin{equation}
\bar{{\cal U}}^{(e)}(t)=tr_{B}\{{\cal U}^{(e)}(t)\},\label{eq:Ue_red}\end{equation}
contains information about the evolution of the RDM only.

\subsection{Non-linear spectroscopy}

In non-linear spectroscopy, coherent laser light is used to investigate
the dynamics of molecular systems by applying special sequences of
pulses. Some pulses act to induce non-equilibrium dynamics (pump),
and other pulses act to monitor (probe) the evolution after the pump.
One of the most advanced of these methods, coherent 2D-ES \cite{Jonas2003a,ChoBook}
measures the response of a system to three pulses traveling in different
directions $\bm{k}_{1}$, $\bm{k}_{2}$ and $\bm{k}_{3}$. The detection
is arranged in such a way (measuring in the direction $-\bm{k}_{1}+\bm{k}_{2}+\bm{k}_{3}$)
that the signal is predominantly of the third order, with contributions
of one order per pulse \cite{MukamelBook}. Let us denote delays between
the first two pulses by $\tau$ and the delay between the second and
the third pulse by $T$. If the pulses are ideally short, the signal
is composed of two kinds of contribution. First, contribution that
involves population of the excited state corresponding to the density
operator \begin{equation}
W_{\bm{k}_{2},\bm{k}_{1}}^{(e)}(t,\tau)=|\psi_{e}^{(\bm{k}_{2})}(t)\rangle\langle\psi_{e}^{(\bm{k}_{1})}(t+\tau)|,\label{eq:We_1}\end{equation}
and second, contribution that involves evolution in the ground state
\begin{equation}
W_{\bm{k}_{2},\bm{k}_{1}}^{(g)}(t,\tau)=|\Psi_{B}\rangle|g\rangle\langle\psi_{g}^{(\bm{k}_{2},\bm{k}_{1})}(t,t+\tau)|.\label{eq:Wg_1}\end{equation}
Here, we denote the pulses acting on the state vector by their corresponding
wave vector in the upper index, and the excited state or ground state
bands by the lower index $g$ and $e$, respectively. For these statistical
operators we can define evolution superoperators ${\cal U}^{(e)}(t)$,
${\cal U}^{(g)}(t)$, ${\cal U}^{(eg)}(t)$ and $U^{(ge)}(t)$ in
analogy with Eqs. (\ref{eq:P_nm_sup}) and (\ref{eq:sup_elements}),
so that\begin{equation}
W_{\bm{k}_{2},\bm{k}_{1}}^{(e)}(t,\tau)={\cal U}^{(e)}(t){\cal U}^{(ge)}(\tau)|\Psi_{B}\rangle\langle\Psi_{B}|\mu|g\rangle\langle g|\mu,\label{eq:We_2}\end{equation}
and \begin{equation}
W_{\bm{k}_{2},\bm{k}_{1}}^{(g)}(t,\tau)={\cal U}^{(g)}(t){\cal U}^{(ge)}(\tau)|\Psi_{B}\rangle\langle\Psi_{B}|\mu^{2}|g\rangle\langle g|.\label{eq:Wg_2}\end{equation}
The superoperator ${\cal U}^{(eg)}(t)$ is the evolution superoperator
of a coherence projector $\sum_{n}|e_{n}\rangle\langle g|$ and analogically
for ${\cal U}^{(ge)}(t)$. After a delay $T$ the third ultrafast
pulse is applied and the non-linear signal is recorded. The signal
corresponds to indirectly to non-linear polarization of the sample,
and is usually measured in frequency domain\[
E^{(3)}(\omega,T,\tau)\approx i\ tr\{\mu{\cal U}^{(eg)}(\omega)\mu\]
\begin{equation}
\times({\cal U}^{(e)}(T){\cal U}^{(ge)}(\tau)\mu\rho_{g}\mu+{\cal U}^{(g)}(T){\cal U}^{(ge)}(\tau)\mu^{2}\rho_{g})W_{eq}\}.\label{eq:E3}\end{equation}
Here, we denoted\begin{equation}
{\cal U}^{(eg)}(\omega)=\int\limits _{0}^{\infty}dt\ e^{i\omega t}{\cal U}^{(eg)}(t),\label{eq:U_omega}\end{equation}
and $\rho_{g}=|g\rangle\langle g|$. In 2D coherent spectroscopy,
the signal is in addition Fourier transformed along the time delay
$\tau$, so that the spectrum is defined as\begin{equation}
S_{2D}(\omega_{t},T,\omega_{\tau})=\int\limits _{-\infty}^{\infty}d\tau\ e^{-i\omega_{\tau}\tau}E^{(3)}(\omega_{t},T,\tau).\label{eq:2D_spect}\end{equation}
The spectrum defined in this way has a suitable interpretation of
an absorption -- absorption and absorption -- stimulated emission
correlation plot, with a different waiting times $T$ between the
two events. The 2D spectrum is in practice measured with finite pulses,
and the measured time domain signal is thus a triple convolution of
the responses to a delta pulse excitation, with the actual finite
pulses \cite{Brixner2005b}. 

From this rough sketch of the principles and the information content
of the coherent 2D spectroscopy it should be clear that 2D spectroscopy
is aimed at disentangling the dynamics of the system during the time
delay $T$. In the so-called Markov approximation, when the dynamics
in time intervals $\tau$, $T$ and $t$ is assumed separable, and
the bath is assumed stationary, the ground state evolution during
interval $T$ can be neglected. Then 2D measurement essentially accesses
the reduced evolution superoperator, Eq. (\ref{eq:Ue_red}) and possibly
also the more general superoperator \begin{equation}
\bar{{\cal U}}^{(e)}(t,\tau)=tr_{B}\{{\cal U}^{(e)}(t)U^{(ge)}(\tau)W_{eq}\}.\label{eq:Ue_ttau}\end{equation}
We will show below that this superoperator, together with the light
properties, determines the way in which the molecule is excited in
a general case, even at illumination by natural light.

\section{Excitation by Light\label{sec:Excitation-by-Thermal}}

In order to account for general light properties we will consider
the problem fully quantum mechanically, and assume only deterministic
evolution of the system wavefunction. The Hamiltonian of the system
reads\[
H=H_{M}+H_{B}+H_{R}+H_{S}\]
\begin{equation}
+H_{M-B}+H_{M-R}+H_{M-S}+H_{B-S}+H_{B-R}+H_{R-S}.\label{eq:Ham_compl}\end{equation}
We have divided the system into a molecule ($H_{M}$), its environment
or bath ($H_{B}$), the radiation ($H_{R}$) and the light emitting
body (LEB) which produces it, e.g. Sun or laser medium ($H_{S}$).
It seems reasonable to neglect a direct interaction between the molecule
(together with its environment) and the molecules of the LEB. Consequently,
the terms $H_{M-S}$ and $H_{B-S}$ can be disregarded. To make the
treatment simpler we can also neglect the interaction between radiation
and the molecular environment, $H_{B-R}$. The assumption is that
the energy of the molecular transition that is used to harvest light
for photosynthetic purposes is much larger than any of the transitions
in this environment and the two regions of the light spectrum can
thus be treated separately. One can also assume that the part of radiation
spectrum which would interact with the bath is simply filtered out,
and the environment is kept at certain temperature by other means.

\subsection{Radiation entangled with the light emitting body}

An important special case is the one in which the radiation and the
LEB is in equilibrium with each other so that the radiation is described
by the canonical equilibrium density matrix\begin{equation}
W_{R}^{(eq)}=\sum_{\lambda\bm{q}}\frac{e^{-\frac{N_{\lambda\bm{q}}\hbar\omega_{\bm{q}}}{k_{B}T}}}{Z_{\lambda\bm{q}}}|N_{\lambda\bm{q}}\rangle\langle N_{\lambda\bm{q}}|.\label{eq:R_eq}\end{equation}
Here, $|N_{\lambda\bm{q}}\rangle$ is the $N$-photon state of the
radiation mode with polarization vector $\bm{e}_{\lambda}$ and wave-vector
$\bm{q}$. As we have already noted above, the statistical concept
of density matrix will be replaced here with the concept of entangled
states, so that we can describe the whole system by its state vector.
Thus, we introduce a state vector \begin{equation}
|\Xi(t)\rangle=\sum_{\lambda\bm{q}}\sum_{N_{\lambda\bm{q}}}c_{N\lambda\bm{q}}(t)|N_{\lambda\bm{q}}\rangle|\phi_{N\lambda\bm{q}}(t)\rangle,\label{eq:Xi}\end{equation}
in which the light is fully entangled with the states $|\phi_{N\lambda\bm{q}}(t)\rangle$
of the LEB . The LEB states have to fulfill the condition \begin{equation}
\langle\phi_{N\lambda\bm{q}}(t)|\phi_{N^{\prime}\lambda^{\prime}\bm{q}^{\prime}}(t)\rangle=\delta_{\lambda\lambda^{\prime}}\delta_{\bm{q}\bm{q}^{\prime}}\delta_{N_{\lambda\bm{q}}N_{\lambda\bm{q}}^{\prime}},\label{eq:condition_phi}\end{equation}
so that when the total density matrix of the LEB and radiation is
averaged over the states of the body, we obtain Eq. (\ref{eq:R_eq}).
$W_{R}^{(eq)}$ is recovered provided that \begin{equation}
|c_{N\lambda\bm{q}}(t)|^{2}=\frac{e^{-\frac{N_{\lambda\bm{q}}\hbar\omega_{\bm{q}}}{k_{B}T}}}{Z_{\lambda\bm{q}}}.\label{eq:coeffs_c}\end{equation}

In the absence of the light absorbing body, the evolution of the state
$|\Xi(t)\rangle$ is governed by the Hamiltonian\begin{equation}
H_{L}=H_{R}+H_{S}+H_{R-S},\label{eq:HL}\end{equation}
and \[
U_{L}(t)=\Theta(t-t_{0})\]
\begin{equation}
\times\exp\left\{ -\frac{i}{\hbar}(H_{S}+H_{R}+H_{R-S})(t-t_{0})\right\} \label{eq:UL}\end{equation}
is the corresponding evolution operator.

\subsection{Equation of motion}

For the subsequent treatment of the system dynamics, we introduce
the interaction picture with respect to Hamiltonian operators $H_{M}$,
$H_{B}$ and $H_{L}$,\begin{equation}
H^{(I)}(t)=H_{M-B}(t)+H_{M-R}(t),\label{eq:H_I}\end{equation}
where\begin{equation}
H_{M-B}(t)=U_{M}^{\dagger}(t)U_{B}^{\dagger}(t)H_{M-B}U_{B}(t)U_{M}(t),\label{eq:HMBt}\end{equation}
and \begin{equation}
H_{M-R}(t)=U_{M}^{\dagger}(t)U_{L}^{\dagger}(t)H_{M-R}U_{L}(t)U_{M}(t).\label{eq:HMRt}\end{equation}
Equation of motion for the total state vector in the interaction picture
\begin{equation}
|\Psi^{(I)}(t)\rangle=U_{M}^{\dagger}(t)U_{B}^{\dagger}(t)U_{L}^{\dagger}(t)|\Psi(t)\rangle\label{eq:I_picture}\end{equation}
 thus reads\[
\frac{\partial}{\partial t}|\Psi^{(I)}(t)\rangle=\]
\begin{equation}
-\frac{i}{\hbar}\left(H_{M-B}(t)+H_{M-R}(t)\right)|\Psi^{(I)}(t)\rangle.\label{eq:Sch_I}\end{equation}
The solution of Eq. (\ref{eq:Sch_I}) can be found formally as\[
|\Psi^{(I)}(t)\rangle=\exp_{+}\Big\{-\frac{i}{\hbar}\int\limits _{t_{0}}^{t}d\tau(H_{M-B}(\tau)\]
\begin{equation}
+H_{M-R}(\tau))\Big\}|\Psi_{0}\rangle.\label{eq:solution_Psi_I}\end{equation}
We will assume that the system is initially in the state $|\Psi_{0}\rangle$
of Eq. (\ref{eq:state_ini}). With this choice we have\begin{equation}
H_{M-B}(t)|\Psi_{0}\rangle=0.\label{eq:HMB_on_Psi_0}\end{equation}
Further in this paper, we will assume weak interaction with the radiation,
so that it can be described by linear theory. Thus, we need to collect
all terms in the expansion of Eq. (\ref{eq:solution_Psi_I}) which
include one occurrence of $H_{M-R}(t)$. Thanks to Eq. (\ref{eq:HMB_on_Psi_0}),
however, all terms where $H_{M-R}(t)$ is not on the far right of
the expression are equal to zero. Eq. (\ref{eq:solution_Psi_I}) therefore
simplifies into a series\[
|\Psi^{(I)}(t)\rangle=|\Psi_{0}\rangle-\frac{i}{\hbar}\int\limits _{t_{0}}^{t}d\tau H_{M-R}(\tau)|\Psi_{0}\rangle\]
\begin{equation}
-\frac{1}{\hbar^{2}}\int\limits _{t_{0}}^{t}d\tau\int\limits _{t_{0}}^{t}d\tau^{\prime}H_{M-B}(\tau)H_{M-R}(\tau^{\prime})|\Psi_{0}\rangle+\dots\ .\label{eq:sol_series}\end{equation}
Now we introduce a projector ${\cal P}_{e}$ that excludes the excitonic
ground state $|g\rangle$\begin{equation}
{\cal P}_{e}=\sum_{n}|e_{n}\rangle\langle e_{n}|.\label{eq:P_e}\end{equation}
Applying this projector to Eq. (\ref{eq:sol_series}) has only the
effect of eliminating the first term of the series. Introducing abbreviations
\begin{equation}
|S(t)\rangle=-\frac{i}{\hbar}\int\limits _{t_{0}}^{t}d\tau H_{M-R}(\tau)|\Psi_{0}\rangle\label{eq:Source_state}\end{equation}
and\begin{equation}
|\Psi_{e}^{(I)}(t)\rangle={\cal P}_{e}|\Psi^{(I)}(t)\rangle,\label{eq:Psi_I_e}\end{equation}
we can write\[
|\Psi_{e}^{(I)}(t)\rangle=|S(t)\rangle-\frac{i}{\hbar}\int\limits _{t_{0}}^{t}d\tau H_{M-B}(\tau)|S(\tau)\rangle\]
\begin{equation}
-\frac{1}{\hbar^{2}}\int\limits _{t_{0}}^{t}d\tau\int\limits _{t_{0}}^{\tau}d\tau^{\prime}H_{M-B}(\tau)H_{M-B}(\tau^{\prime})|S(\tau^{\prime})\rangle+\dots\ .\label{eq:P_e_series}\end{equation}
It is possible to verify easily that this series is a solution of
the equation \begin{equation}
\frac{\partial}{\partial t}|\Psi_{e}^{(I)}(t)\rangle=-\frac{i}{\hbar}H_{M-B}(t)|\Psi_{e}^{(I)}(t)\rangle+\frac{\partial}{\partial t}|S(t)\rangle,\label{eq:source_eq}\end{equation}
with initial condition $|\Psi_{e}^{(I)}(t_{0})\rangle=0$.

\subsection{Pumping source term}

Eq. (\ref{eq:source_eq}) is an equation of motion for the excited
states of an excitonic aggregate pumped by a source term \begin{equation}
|S^{\prime}(t)\rangle=\frac{\partial}{\partial t}|S(t)\rangle=-\frac{i}{\hbar}H_{M-R}(t)|\Psi_{0}\rangle.\label{eq:S_prime}\end{equation}
Hamiltonian $H_{M-R}$ will be assumed in the dipole approximation,
i.e.\begin{equation}
H_{M-R}\approx-\bm{\mu}\cdot\bm{E}_{T}(\bm{r}),\label{eq:H_MR_form}\end{equation}
where $\bm{\mu}$ is the transition dipole moment operator of the
aggregate\begin{equation}
\bm{\mu}=\sum_{n}\bm{d}_{n}|e_{n}\rangle\langle g|+h.c.,\label{eq:mu_def}\end{equation}
and $\bm{E}_{T}$ is the operator of the (transversal) electric field
of the radiation\[
\bm{E}_{T}(\bm{r})=-i\sum_{\lambda\bm{q}}\Big(\bm{e}_{\lambda-\bm{q}}f_{-\bm{q}}(\bm{r})a_{\lambda\bm{q}}^{\dagger}\]
\begin{equation}
-\bm{e}_{\lambda\bm{q}}f_{\bm{q}}(\bm{r})a_{\lambda\bm{q}}\Big),\label{eq:E_T_def}\end{equation}
with \begin{equation}
f_{\bm{q}}(\bm{r})=\sqrt{\frac{\hbar\omega_{\bm{q}}}{2\epsilon_{0}\Omega}}e^{i\bm{q}\cdot\bm{r}}.\label{eq:flq}\end{equation}
Here, $\Omega$ is a quantization volume. 

We consider a molecule much smaller than the wavelength of the light,
so that $e^{i\bm{q}\cdot\bm{r}}$ is constant in the volume of the
molecule. The origin of the coordinates can thus be conveniently put
into the molecule yielding $e^{i\bm{q}\cdot\bm{r}}\approx1$. The
interaction Hamiltonian in Eq. (\ref{eq:S_prime}) then reads\[
H_{M-R}(t)=i\sum_{\lambda\bm{q}}\mu_{\lambda\bm{q}}(t)f_{\lambda-\bm{q}}(0)a_{\lambda\bm{q}}^{\dagger}(t)\]
\begin{equation}
-i\mu_{\lambda\bm{q}}(t)f_{\lambda\bm{q}}(0)a_{\lambda\bm{q}}(t),\label{eq:HMR_in_a}\end{equation}
where the creation and annihilation operators of the field are in
the interaction picture with respect to Hamiltonian $H_{L}$, i.e.\begin{equation}
a_{\lambda\bm{q}}^{\dagger}(t)=U_{L}^{\dagger}(t)a_{\lambda\bm{q}}^{\dagger}U_{L}(t),\label{eq:adagger_t}\end{equation}
\begin{equation}
a_{\lambda\bm{q}}(t)=U_{L}^{\dagger}(t)a_{\lambda\bm{q}}U_{L}(t).\label{eq:a_t}\end{equation}
The transition dipole moment operator projected on the polarization
vector of a mode $\lambda\bm{q}$ appears in the interaction picture
with respect to Hamiltonian $H_{M}$,\begin{equation}
\mu_{\lambda\bm{q}}(t)=U_{M}^{\dagger}(t)\bm{\mu}\cdot\bm{e}_{\lambda\bm{q}}U_{M}(t).\label{eq:mu_lq}\end{equation}
The evolution operator $U_{L}(t)$, Eq. (\ref{eq:UL}), can be rewritten
as\begin{equation}
U_{L}(t)=U_{S}(t)U_{R}(t){\cal U}_{R-S}(t),\label{eq:UL_SR}\end{equation}
where\[
U_{R-S}(t)=\Theta(t-t_{0})\exp_{+}\Big\{-\frac{i}{\hbar}\int\limits _{t_{0}}^{t}d\tau U_{S}^{\dagger}(\tau)U_{R}^{\dagger}(\tau)\]
 \begin{equation}
\times H_{R-S}U_{R}(\tau)U_{S}(\tau)\Big\}.\label{eq:calU_RS}\end{equation}
Since Hamiltonian $H_{S}$ commutes with the radiation operators and
\begin{equation}
U_{R}^{\dagger}(t)a_{\lambda\bm{q}}U_{R}(t)=e^{-i\omega_{\bm{q}}t}a_{\lambda\bm{q}},\label{eq:UaU}\end{equation}
we have\begin{equation}
a_{\lambda\bm{q}}^{\dagger}(t)=\tilde{a}_{\lambda\bm{q}}^{\dagger}(t)e^{i\omega_{\bm{q}}t},\label{eq:adag}\end{equation}
and \begin{equation}
a_{\lambda\bm{q}}(t)=\tilde{a}_{\lambda\bm{q}}(t)e^{-i\omega_{\bm{q}}t}.\label{eq:a}\end{equation}
Here, we introduced slow oscillating envelops \begin{equation}
\tilde{a}_{\lambda\bm{q}}^{\dagger}(t)=U_{R-S}^{\dagger}(t)a_{\lambda\bm{q}}^{\dagger}U_{R-S}(t),\label{eq:adag_til}\end{equation}
and\begin{equation}
\tilde{a}_{\lambda\bm{q}}(t)=U_{R-S}^{\dagger}(t)a_{\lambda\bm{q}}U_{R-S}(t).\label{eq:a_til}\end{equation}
Inserting these expressions into Eq. (\ref{eq:HMR_in_a}) we can distinguish
two terms associated with the transition from the ground state $|g\rangle$
to an excited state $|e_{a}\rangle$ with respective phase factors
$e^{i(\omega_{ag}-\omega_{\bm{q}})t}$ and $e^{i(\omega_{ag}+\omega_{\bm{q}})t}$.
While the first one will lead to a resonance excitation around $\omega_{\bm{q}}\approx\omega_{ag}$,
the later term is oscillating fast and will therefore contribute very
little compared to the former one. Thus we drop the fast oscillating
part, and obtain the source term in the form\[
|S^{\prime}(t)\rangle=\frac{1}{\hbar}\sum_{\lambda\bm{q}}\mu_{\lambda\bm{q}}(t)f_{\lambda\bm{q}}\]
\begin{equation}
\times a_{\lambda\bm{q}}(t)|g\rangle|\Phi_{B}\rangle|\Xi_{0}\rangle.\label{eq:S_prime_derived}\end{equation}
Using this form of the source term, we can find state into which the
molecule is weakly driven by any type of light.

\subsection{Excited state dynamics under pumping}

So far we have treated the problem systematically using the wavefunction
approach. The time evolution of the system wavefunction is governed
by Eq. (\ref{eq:Sch_I}). To find the probabilities of creating population
on and coherence between certain excitonic levels $|e_{a}\rangle$
we solve Eq. (\ref{eq:Sch_I}) formally,\begin{equation}
|\Psi_{e}^{(I)}(t)\rangle=\int\limits _{t_{0}}^{t}d\tau U_{M-B}(t-\tau)|S^{\prime}(\tau)\rangle.\label{eq:solution_psi_e}\end{equation}
Here, we used the fact that $|\Psi_{e}(t_{0})\rangle=0$. Now let
us evaluate matrix element $P_{ab}(t)=\langle\Psi_{e}(t)|{\cal P}_{ab}|\Psi_{e}(t)\rangle$
of a projector \begin{equation}
{\cal P}_{ab}=|e_{a}\rangle\langle e_{b}|,\label{eq:proj_nm}\end{equation}
 which gives the probability of finding the molecule in state $|e_{a}\rangle$
if $a=b$, or characterizes the amount of coherence between states
$|e_{a}\rangle$ and $|e_{b}\rangle$ if $a\neq b$. Note that we
have removed the interaction picture, Eq. (\ref{eq:I_picture}). We
have\[
P_{ab}(t)=\frac{1}{\hbar^{2}}\int\limits _{t_{0}}^{t}d\tau\int\limits _{t_{0}}^{t}d\tau^{\prime}\sum_{\lambda\bm{q},\lambda^{\prime}\bm{q}^{\prime}}f_{\lambda\bm{q}}(f_{\lambda^{\prime}\bm{q}^{\prime}})^{*}\]
\[
\times\langle\Xi_{0}|\tilde{a}_{\lambda\bm{q}}^{\dagger}(\tau)\tilde{a}_{\lambda^{\prime}\bm{q}^{\prime}}(\tau^{\prime})|\Xi_{0}\rangle\]
\begin{equation}
\times\langle e_{b}|\bar{{\cal U}}^{(e)}(t-\tau,\tau-\tau^{\prime})\rho_{\lambda\bm{q},\lambda^{\prime}\bm{q}^{\prime}}^{0}|e_{a}\rangle,\label{eq:Pab_general}\end{equation}
where the evolution superoperator $\bar{{\cal U}}^{(e)}(t-\tau,\tau-\tau^{\prime})$
has been defined in Eq. (\ref{eq:Ue_ttau}). 

In Eq. (\ref{eq:Pab_general}), the light is represented by a first
order correlation function\[
I_{\lambda\bm{q},\lambda^{\prime}\bm{q}}^{(1)}(\tau,\tau^{\prime})=(f_{\lambda\bm{q}}(f_{\lambda^{\prime}\bm{q}^{\prime}})^{*})^{-1}\]
 \begin{equation}
\times\langle\Xi_{0}|\tilde{a}_{\lambda\bm{q}}^{\dagger}(\tau)\tilde{a}_{\lambda^{\prime}\bm{q}^{\prime}}(\tau^{\prime})|\Xi_{0}\rangle\label{eq:I1_def}\end{equation}
 (see e. g. Ref. \cite{LoundonBook}), which comprises all its relevant
properties. We also denoted \begin{equation}
\rho_{\lambda\bm{q},\lambda^{\prime}\bm{q}^{\prime}}^{0}=\frac{1}{\hbar^{2}}\mu_{\lambda\bm{q}}|g\rangle\langle g|\mu_{\lambda^{\prime}\bm{q}^{\prime}}.\label{eq:rho_ini}\end{equation}
The quantities $P_{ab}(t)$ are the matrix elements of the RDM ($P_{ab}(t)=\langle e_{b}|\rho(t)|e_{a}\rangle$)
of the system which reads\[
\rho(t)=\int\limits _{t_{0}}^{t}d\tau\int\limits _{t_{0}}^{t}d\tau^{\prime}\bar{{\cal U}}^{(e)}(t-\tau,\tau-\tau^{\prime})\]
\begin{equation}
\times\sum_{\lambda\bm{q},\lambda^{\prime}\bm{q}^{\prime}}\rho_{\lambda\bm{q},\lambda^{\prime}\bm{q}^{\prime}}^{0}I_{\lambda\bm{q},\lambda^{\prime}\bm{q}}^{(1)}(\tau,\tau^{\prime}).\label{eq:rho_final}\end{equation}
For a weakly driven system, Eq. (\ref{eq:rho_final}) has a very wide
range of applicability. We will discuss its application to thermal
light and pulsed coherent light in the following section.

\section{Discussion\label{sec:Discussion}}

Thorough discussion of excitation dynamics of molecular systems excited
by incoherent light was made in Ref. \cite{Brumer}. Molecular systems
were considered without the bath effect which is however significant
for light harvesting. Eq. (\ref{eq:rho_final}) contains reduced evolution
superoperator of the molecular system so that the state of the system
created by the incident light depends on its reduced dynamics. It
is not possible to consider a general case of such dynamics analytically,
and we will therefore commit ourselves to some simple cases. 

In so called secular and Markov approximations (see e.g. Ref. \cite{MayKuehnBook})
matrix elements of the evolution superoperator governing the coherences
take a very simple form. First, it is possible to separate the two
time arguments in the superoperator $\bar{{\cal U}}^{(e)}(t,\tau)$
so that \begin{equation}
\bar{{\cal U}}^{(e)}(t,\tau)=\bar{{\cal U}}^{(e)}(t)\bar{{\cal U}}^{(eg)}(\tau).\label{eq:Markov_U}\end{equation}
Since each coherence is independent of the population dynamics and
of other coherences, the one-argument superoperator elements read\begin{equation}
\bar{{\cal U}}_{abab}^{(e)}(t)=e^{-i\omega_{ab}t-(\Gamma_{a}+\Gamma_{b})t},\label{eq:Uabab}\end{equation}
and\begin{equation}
\bar{{\cal U}}_{agag}^{(eg)}(t)=e^{-i\omega_{ag}t-\Gamma_{a}t}.\label{eq:Uagag}\end{equation}
Here the dephasing rate \begin{equation}
\Gamma_{a}=\gamma_{p}+\frac{1}{2}K_{a}\label{eq:Gamma_ab}\end{equation}
comprises the pure dephasing rate $\gamma_{p}$ and the rate $K_{a}$
of depopulation, i.e. the sum of transition rates from state $|e_{a}\rangle$
to other states. A simplified treatment of the populations is possible
for the states that are only depopulated, i.e. no contributions to
the population can be attributed to the transfer from other levels.
They are found at the top of the energetic funnel of the antenna.
For these states we have\begin{equation}
\bar{{\cal U}}_{aaaa}^{(e)}(t)=e^{-K_{a}t}.\label{eq:Uaaaa}\end{equation}
Eqs. (\ref{eq:Uabab}) to (\ref{eq:Uaaaa}) neglect all coherence
transfer effects, as well as possible coupling between the dynamics
of population and coherence.

\subsection{Excitation of coherences by thermal light}

For an equilibrium thermal light the correlation function $I_{\lambda\bm{q},\lambda^{\prime}\bm{q}^{\prime}}^{(1)}(\tau,\tau^{\prime})$
depends only on the difference of the times $\tau$ and $\tau^{\prime}$.
As discussed above, $|\Xi_{0}\rangle$ represents the equilibrium
of the system described by Hamiltonian $H_{L}$. The equilibrium density
matrix is stationary, i.e.\begin{equation}
U_{L}(t)|\Xi_{0}\rangle\langle\Xi_{0}|U_{L}^{\dagger}(t)=|\Xi_{0}\rangle\langle\Xi_{0}|,\label{eq:dens_eq}\end{equation}
 so we can write\[
I_{\lambda\bm{q},\lambda^{\prime}\bm{q}}^{(1)}(\tau,\tau^{\prime})=|f_{\lambda\bm{q}}|^{-2}\langle\Xi_{0}|\tilde{a}_{\lambda\bm{q}}^{\dagger}(\tau-\tau^{\prime})\tilde{a}_{\lambda\bm{q}}(0)|\Xi_{0}\rangle\]
\[
\times e^{i\omega_{\bm{q}}(\tau-\tau^{\prime})}\delta_{\lambda\lambda^{\prime}}\delta_{\bm{q}\bm{q}^{\prime}}\]
\begin{equation}
\equiv|f_{\lambda\bm{q}}|^{-2}\tilde{I}_{\lambda\bm{q}}(\tau-\tau^{\prime})\delta_{\lambda\lambda^{\prime}}\delta_{\bm{q}\bm{q}^{\prime}}.\label{eq:n_def}\end{equation}
It can be shown that \begin{equation}
\tilde{I}_{\lambda\bm{q}}(-t)=\tilde{I}_{\lambda\bm{q}}^{*}(t).\label{eq:n_sym}\end{equation}
Assuming some simple form of a light correlation function, e.g. \begin{equation}
\tilde{I}_{\lambda\bm{q}}(t)=I_{\lambda\bm{q}}^{0}e^{-\frac{|t|}{\tau_{d}}+i\omega_{\bm{q}}t},\label{eq:I_collisional}\end{equation}
we obtain for the populations\[
\rho_{aa}(t)=2Re\sum_{\lambda\bm{q}}I_{\lambda\bm{q}}^{0}[\rho_{\lambda\bm{q}}]_{aa}\int\limits _{t_{0}}^{t}d\tau\int\limits _{t_{0}}^{\tau}d\tau^{\prime}e^{-K_{a}(t-\tau)}\]
\begin{equation}
\times e^{-\Gamma_{a}(\tau-\tau^{\prime})-\frac{\tau-\tau^{\prime}}{\tau_{d}}-i(\omega_{ag}-\omega_{\bm{q}})(\tau-\tau^{\prime})}\label{eq:pop_integral_form}\end{equation}
Here, $[\rho]_{ab}\equiv\langle e_{a}|\rho|e_{b}\rangle$. We utilized
Eq. (\ref{eq:I_collisional}) and the fact that, by definition (see
Eqs. (\ref{eq:I1_def}) and (\ref{eq:rho_ini})), the time $\tau$
corresponds to the action of the dipole moment operator from left,
whereas time $\tau^{\prime}$ corresponds to the same action from
the right. At long times $t-t_{0}\rightarrow\infty$ this yields \begin{equation}
\rho_{aa}^{\infty}=\sum_{\lambda\bm{q}}\frac{2}{K_{a}}\frac{(\Gamma_{a}+\frac{1}{\tau_{d}})I_{\lambda\bm{q}}^{0}[\rho_{\lambda\bm{q}}^{0}]_{aa}}{(\omega_{ag}-\omega_{\bm{q}})^{2}+(\Gamma_{a}+\frac{1}{\tau_{d}})^{2}}.\label{eq:rho_aa_inf}\end{equation}
However, neglecting the influence of environment as in Ref. \cite{Brumer}
yields\begin{equation}
\rho_{aa}^{long}(t-t_{0})=\sum_{\lambda\bm{q}}\frac{\tau_{d}^{-1}2I_{\lambda\bm{q}}^{0}[\rho_{\lambda\bm{q}}^{0}]_{aa}(t-t_{0})}{(\omega_{ag}-\omega_{\bm{q}})^{2}+\frac{1}{\tau_{d}^{2}}},\label{eq:rho_aa_delta}\end{equation}
which grows linearly with time.

For coherences we have\[
\rho_{ab}^{\infty}=2\sum_{\lambda\bm{q}}I_{\lambda\bm{q}}^{0}[\rho_{\lambda\bm{q}}^{0}]_{ab}\frac{1}{i\omega_{ab}+(\Gamma_{a}+\Gamma_{b})}\]
\[
\times\Big[\frac{1}{i(\omega_{ag}-\omega_{\bm{q}})+\Gamma_{a}+\frac{1}{\tau_{b}}}\]
\begin{equation}
+\frac{1}{-i(\omega_{bg}-\omega_{\bm{q}})+\Gamma_{b}+\frac{1}{\tau_{b}}}\Big],\label{eq:rho_ab_inf}\end{equation}
which turns into Eq. (\ref{eq:rho_aa_inf}) for $a=b$ (with additional
assumption $K_{a}=2\Gamma_{a}$). In case of no dephasing, the first
fraction in Eq. (\ref{eq:rho_ab_inf}) yields a delta function $\delta(\omega_{ab})$
\cite{Brumer}. Thus, for slow or non-existent relaxation due to interaction
with environment, the system is excited predominantly into a state
represented by a diagonal RDM, as all coherence terms are negligible
compared to the linearly growing population. For fast relaxation,
the coherences may be of the same order of magnitude as the populations.

The case of very fast relaxation is particularly interesting. It was
suggested previously that coherent dynamics can be relevant for the
\emph{in vivo} case, because the fluctuating light from the Sun corresponds
to a train of ultrafast spikes \cite{Cheng-Review}. The relaxation
of the antenna must be in such a case fast enough to prevent averaging
over many such spikes. Eqs. (\ref{eq:rho_aa_inf}) and (\ref{eq:rho_ab_inf})
with large $K_{a}$ describe just such a situation. The RDM created
by incoherent light resembles in certain sense the one created by
ultrafast pulses; it represents a linear combination of excitons.
The coherences in Eq. (\ref{eq:rho_ab_inf}) are however static at
long times.

In our demonstration we concentrated on a simple model assuming both
Markov and secular approximations to be valid. The presence or absence
of coherences has no significance in such a case, and more involved
theories of the RDM dynamics \cite{Ishizaki1,Ishizaki2,Olsina2010a}
have to be used to investigate the role of coherences in energy transfer
processes by Eq. (\ref{eq:rho_final}).

\subsection{Coherent pulsed light}

In derivation of Eq. (\ref{eq:rho_final}) we assumed certain initial
state $|\Xi_{0}\rangle$ of the system composed of the light and its
source. The condition that the light is in a stationary state, fully
entangled with its source, has only been used to simplify the correlation
function $I_{\lambda\bm{q},\lambda^{\prime}\bm{q}^{\prime}}^{(1)}(\tau,\tau^{\prime})$
for the case of the thermal light. In a general case $|\Xi_{0}\rangle$
will not represent an equilibrium state. It can indeed describe even
systems such as a laser producing coherent Gaussian light pulses with
some carrier frequency $\omega_{0}$ and a width parameter $\Delta$.
 If we in addition assume that the light is described by a single
polarization, and that we consider the dynamics after one such pulse
centered at time $t=\tau_{0}$, the light is described as\begin{equation}
\sum_{\lambda\bm{q},\lambda^{\prime}\bm{q}^{\prime}}I_{\lambda\bm{q}\lambda^{\prime}\bm{q}^{\prime}}^{(1)}(\tau,\tau^{\prime})=I_{0}e^{-\frac{(\tau-\tau_{0})^{2}}{\Delta^{2}}-\frac{(\tau^{\prime}-\tau_{0})^{2}}{\Delta^{2}}}.\label{eq:Gauss_exc}\end{equation}
 Coherence element created by such light reads\[
\rho_{ba}(t)=e^{i\omega_{ab}t}\int\limits _{t_{0}}^{t}d\tau\int\limits _{t_{0}}^{t}d\tau^{\prime}e^{-i(\omega_{ag}-\omega_{\bm{q}})\tau}e^{i(\omega_{bg}-\omega_{\bm{q}})\tau^{\prime}}\]
\begin{equation}
\times e^{-\Gamma_{a}(t-\tau)-\Gamma_{b}(t-\tau^{\prime})}I_{0}e^{-\frac{(\tau-\tau_{0})^{2}}{\Delta^{2}}-\frac{(\tau^{\prime}-\tau_{0})^{2}}{\Delta^{2}}}\rho_{ba}^{0},\label{eq:Pab_model_coh}\end{equation}
where $\rho_{ba}^{0}=\frac{1}{\hbar^{2}}\langle e_{b}|\mu|g\rangle\langle g|\mu|e_{a}\rangle$.
In the limit of ultrashort pulses when $e^{-\frac{(\tau-\tau_{0})^{2}}{\Delta^{2}}}\rightarrow\alpha\delta(\tau-\tau_{0})$
the pulse creates a pure state at $\tau_{0}$, which then dephases
as\[
\rho_{ba}(t)=\Theta(t-\tau_{0})e^{-(\Gamma_{a}+\Gamma_{b})(t-\tau_{0})}\]
 \begin{equation}
\times e^{i\omega_{ab}(t-\tau_{0})}\rho_{ba}^{0}I_{0}\alpha^{2}.\label{eq:rho_ba_time}\end{equation}
In case of a finite pulse and no dephasing our results coincides with
those found in Ref. \cite{Brumer}.

\subsection{Mediated excitation}

The major difference between excitation by the thermal light and a
coherence pulse is in the occurrence of a sudden event which populates
a nearly pure state of the excited state band. Clearly, a single molecule
interacting with an ideal continuum of radiation modes in equilibrium
does not experience such sudden events. Rather, its interaction with
light corresponds to a continuous pumping, and the suddenness of the
photon arrival is the consequence of our ability to register only
classical outcomes. In order to register them we have to interact
with the system and become entangled with it. Our experience is that
macroscopic systems interacting with low intensity light can be used
to detect single photons, and certain more or less definite times
can be attributed to their arrivals. Interaction of a photon with
a macroscopic detector yields a temporal localization of the arrival
event. A mesoscopic system may play a role of such a detector (mediator)
that provides its fluctuations to be harvested by dedicated nano sized
antenna. Green photosynthetic bacteria, from which the photosynthetic
complex FMO was isolated, collect light mainly by means of so-called
chlorosoms \cite{Blankenship,Chlorosom}. The chlorosom is a self-assembled
aggregate of $\sim10^{5}$ BChls and carotenoids with very little
protein. The typical dimensions the chlorosom are of the order of
$100$ nm \cite{Chlorosom}. It does not seem to be organized as an
energy funnel \cite{Nozawa94,Psencik04}, and the energy transfer
time between its main body and the base plate to which FMO complexes
are attached is of the order of $120$ ps \cite{Psencik03}, i.e rather
slow. The excitation in such a mesoscopic system may have enough time
to become localized through interaction with the large number of the
systems DOF and arrive at the FMO complex in a particle like, i.e.
also temporally localized fashion.

In this section, we will generalize our result, Eq. (\ref{eq:rho_final}),
for a case when the excitation of the photosynthetic systems occurs
by transfer from another system. We will therefore assume that our
molecule does not interact directly with light, but is pumped in a
similar fashion by another system. The source term, Eq. (\ref{eq:S_prime_derived}),
is then generalized as\[
|S^{\prime}(t)\rangle=\frac{i}{\hbar}A(t)|g\rangle|\Phi_{B}\rangle\]
\begin{equation}
\times\left(\sum_{n}\alpha_{n}(t)|\xi_{n}\rangle|\phi_{n}(t)\rangle\right).\label{eq:S_prime_gene}\end{equation}
Here, $A=\sum_{\alpha,n}|e_{n}\rangle|\xi_{g}\rangle\langle\xi_{\alpha}|\langle g|+h.c.$
is the molecule--mediator interaction Hamiltonian and the time dependence
results from the interaction picture\begin{equation}
A(t)=U_{M}^{\dagger}(t)U_{A}^{\dagger}(t)AU_{A}(t)U_{M}(t).\label{eq:A_op}\end{equation}
We denoted the ground and excited states of the mediator by $|\xi_{g}\rangle$
and $|\xi_{n}\rangle$, respectively. The state of the molecule at
long times is in analogy with Eq. (\ref{eq:rho_final})\[
\rho(t)=\int\limits _{t_{0}}^{t}d\tau\int\limits _{t_{0}}^{t}d\tau^{\prime}{\cal U}^{(e)}(t-\tau,\tau-\tau^{\prime})\sum_{nn^{\prime}}\alpha_{n}^{*}(\tau)\alpha_{n^{\prime}}(\tau)\]
\begin{equation}
\times\langle\xi_{n}|A(\tau)A(\tau^{\prime})|\xi_{n^{\prime}}\rangle\langle\phi_{n}(\tau)|\phi_{n^{\prime}}(\tau^{\prime})\rangle.\label{eq:rho_fina_mosol}\end{equation}
The complicated two-point correlation function in Eq. (\ref{eq:rho_fina_mosol})
results from the pumping of the mediator similarly to the direct pumping
of the molecule in Eq. (\ref{eq:rho_final}). A mesoscopic system
especially when excited will, however, always exhibit fluctuations
which will prevent the correlation function from having a simple smooth
dependence without recurrences. Such recurrences can temporally localize
the excitation events of the molecule. In such an excitation regime,
when coherent dynamics from different excitation times do not interfere,
optimization of the FMO's energy channeling capability for case of
initially coherent states would be an advantage.

\subsection{Outlook }

More research into specific forms of both the light correlation function
for different situation that may occur \emph{in vivo}, and the analogical
interaction of systems like FMO with mesoscopic antennae is clearly
needed.\emph{ }Ultrafast spectroscopic experiments play a pivotal
role in this research by yielding information about the system response
to the light. To conclude on the utility of coherent dynamics for
the function of the photosynthetic system is, however, only possible
by taking into account the properties of light at the natural conditions,
for which the results of this paper provide means. If the coherent
dynamics observed in some photosynthetic chromophore-protein complexes
has a significance for their light-harvesting efficiency, and these
systems evolved to optimize it for their corresponding ecological
situation, it can be expected that the properties of at least some
parts of the photosynthetic machinery would be tuned to the fluctuation
properties of their source of excitation. For plants and some bacteria
this may be the Sun light, others like FMO complexes could be expected
to be tuned to the properties of their associated chlorosoms.

\section{Conclusions\label{sec:Conclustions}}

In this paper we have discussed dynamics of a molecular system subject
to external pumping by a light source. In particular we have considered
excitation by thermal light, by coherent pulsed light and an excitation
through a mesoscopic antenna. With a completely quantum mechanical
treatment, we have derived a general formula which enables us to study
the effect of different light properties on the photo-induced dynamics
of a molecular systems. This formula naturally contains the system--environment
interaction contribution to the excitation process which enters via
appearance of the reduced density matrix dynamics. We show that once
the properties of light are known in terms of a certain two-point
correlation function, the only information needed to reconstruct the
systems dynamics is the reduced evolution superoperator, which is
in principle accessible through ultrafast non-linear spectroscopy.
This conclusion applies to any type of light and makes thus the results
of ultrafast spectroscopic experiments universally relevant. Considering
a direct excitation of a small molecular antenna we found that excitation
of coherences is possible due to overlap of homogeneous line shapes
associated with different excitonic states. These coherences are however
static and correspond to a change of the preferred basis set into
which the system relaxes from the one defined by the bath only, to
the one defined by the action of both the light and the bath. When
an excitation of a photosynthetic complex mediated by a larger, possibly
mesoscopic, system is considered, the complex can harvest fluctuations
originating from the non-equilibrium state of the mediator. Fluctuations
of the mesoscopic system such as chlorosoms may time localize excitation
events of the energy channeling complex, and to excite adjacent energy
channeling complex coherently. It is likely that in such a case the
properties of energy channeling complexes like the well-know Fenna-Mathews-Olson
complex would be specially tuned to the fluctuation properties of
their associated chlorosoms. 
\begin{acknowledgments}
This work was partially supported by the Czech Science Foundation
(GACR) through grant. nr. 206/09/0375, by the Ministry of Education,
Youth and Sports of the Czech Republic through grant KONTAKT ME899
and the research plan MSM0021620835, and by the Agency of the International
Science and Technology in Lithuania.

\bibliographystyle{prsty}
\bibliography{mancal_20100204}

\end{acknowledgments}

\end{document}